\documentstyle[12pt,epsfig,amsmath,amssymb,pstricks,natbib]{article}
\begin{document}

\title{\bf Consistent quantum mechanics admits no mereotopology}

\author{{Chris Fields}\\ \\
{\it 815 East Palace \# 14}\\
{\it Santa Fe, NM 87501 USA}\\ \\
{fieldsres@gmail.com}}
\maketitle
\maketitle

\begin{abstract}
It is standardly assumed in discussions of quantum theory that physical systems can be regarded as having well-defined Hilbert spaces.  It is shown here that a Hilbert space can be consistently partitioned only if its components are assumed not to interact.  The assumption that physical systems have well-defined Hilbert spaces is, therefore, physically unwarranted.
\end{abstract}


\section{Introduction}

The term ``system'' is employed ubiquitously in quantum mechanics.  Formally, a system $\mathbf{S}$ is a collection of physical degrees of freedom that can be specified by a Hilbert space $\mathcal{H}_{\mathbf{S}}$.  Assuming that it makes sense to talk about the Hilbert space $\mathcal{H}_{\mathbf{U}}$ of the universe $\mathbf{U}$ as a whole, a system is a component of a Hilbert-space decomposition or tensor-product structure (TPS) defined on $\mathcal{H}_{\mathbf{U}}$, i.e. $\mathcal{H}_{\mathbf{U}} = \mathcal{H}_{\mathbf{S}} \otimes \mathcal{H}_{\mathbf{E}}$, where $\mathcal{H}_{\mathbf{E}}$ is the Hilbert space of a second system $\mathbf{E}$ that is standardly called the ``environment'' of $\mathbf{S}$.  If the notion of $\mathcal{H}_{\mathbf{U}}$ is well-defined, then clearly for any two systems $\mathbf{S1}$ and $\mathbf{S2}$, there must exist environments $\mathbf{E1}$ and $\mathbf{E2}$ for which $\mathcal{H}_{\mathbf{S1}} \otimes \mathcal{H}_{\mathbf{E1}} = \mathcal{H}_{\mathbf{S2}} \otimes \mathcal{H}_{\mathbf{E2}} = \mathcal{H}_{\mathbf{U}}$.

One can ask, however, whether this formal usage corresponds in a fully-intelligible way to our informal use of the term ``system'' to refer to collections of degrees of freedom being subjected to analysis in the laboratory, whether these collections of degrees of freedom describe macroscopic items of apparatus or ``isolated'' microscopic systems such as pairs of photons in asymmetric Bell states.  In particular, one can ask (1) whether the commonplace laboratory entities that we informally refer to as ``systems'' are systems in the strict sense of being components of well-defined TPSs of $\mathcal{H}_{\mathbf{U}}$, and (2) whether such ``systems'' can be said to ``emerge'' from $\mathcal{H}_{\mathbf{U}}$ by the action of a universal Hamiltonian dynamics $H_{\mathbf{U}}$ that satisfies a Schr\"odinger equation $(\partial / \partial t) | \mathbf{U}\rangle = -(\imath / \hbar) \mathit{H}_{\mathbf{U}} | \mathbf{U}\rangle$.  The non-trivial nature of the latter question has been highlighted by W. Zurek, who remarks that ``it is far from clear how one can define systems given an overall Hilbert space `of everything' and the total Hamiltonian'' (\cite{zurek:98} p. 1794) and in later papers offers as ``axiom(o)'' of quantum mechanics that ``systems exist'' (\cite{zurek:03} p. 746) or ``the Universe consists of systems'' (\cite{zurek:05env} p. 2).  The first question, however, is generally regarded as unproblematic.  It is answered in the affirmative whenever someone implicitly defines a system $\mathbf{S}$ by writing down an expression such as `let $|\mathbf{S}\rangle = \mathit{\sum_{i} \lambda_{i} |s_{i}\rangle}$' that explicitly defines its state $|\mathbf{S}\rangle$ in terms of a collection of Hilbert-space basis vectors $|s_{i}\rangle$, and then proceeds to investigate the system so defined experimentally.

Our informal understanding of systems is fundamentally mereological: systems are conceived as being composed of parts, and their degrees of freedom are conceived as derivative from, or at least emergent from, the degrees of freedom of those parts.  Systems are, moreover, conceived as having boundaries that separate them from other systems.  Hence a magnetic ion trap apparatus, for example, has as identifiable parts a vacuum chamber, an ion source, pumps, magnets, lasers and detectors; it has a well defined ``inside'' that contains the trapped ions and a well-defined ``outside'' that is exposed to the laboratory environment, including the experimenters.  Ions trapped in the ``inside'' are explicitly considered to be distinct systems that can be inserted into, examined within, and extracted from the ion trap; this distinction is reflected in the formal analysis of data collected from ion traps (e.g. \cite{brune:96} where the timecourse of decoherence within an ion trap is analyzed).  What an ion trap apparatus can do - its degrees of freedom - are moreover conceived as entirely dependent on what its various parts can do; if the ion source stops working properly, for example, the ion trap apparatus as a whole stops working properly.  One would, therefore, expect that a self-consistent description of such systems could be provided within the formal framework of mereotopology, i.e. the formal theory of bounded parts and their containing bounded wholes \citep{varzi:94, varzi:96, smith:96, casati:99}.  In particular, one would expect that what are informally regarded as parts of some system $\mathbf{S}$ correspond in an intuitively sensible way to a TPS of $\mathcal{H}_{\mathbf{S}}$; that if $\mathbf{P1}$, $\mathbf{P2}$ ... $\mathbf{PN}$ are parts of $\mathbf{S}$, then $\mathcal{H}_{\mathbf{S}} = \mathcal{H}_{\mathbf{P1}} \otimes \mathcal{H}_{\mathbf{P2}} \otimes ... \otimes \mathcal{H}_{\mathbf{PN}}$.  One might even expect, in the spirit of Zurek's ``axiom(o)'', that a sufficiently detailed dynamical description would, together with the structure of $\mathcal{H}_{\mathbf{U}}$, \textit{require} a unique mereotopology relating systems to their included parts, and hence predict the actual, objective existence a unique set of observable composite objects.  The theoretical project of explaining the ``emergence of classicality'' from the quantum world \citep{schloss:04, schloss:07, landsman:07, wallace:08} can be interpreted as a search for such a unique, dynamically-specified mereotopology. 

It has previously been shown that the dynamics of system-environment decoherence, even when combined with environmental ``witnessing'' and encoding of quantum states \citep{zurek:04, zurek:05, zurek:06, zurek:09}, is insufficient to unambiguously specify the boundaries of macroscopic systems that are regarded by observers as classical \citep{fields:10, fields:11}.  It has also been shown that the ability of observers to obtain classical information about the behavior of collections of quantum degrees of freedom can be consistently described in a boundary-free and hence mereologically-nihilist framework \citep{fields:12a}.  These results call the existence of a unique, dynamically-specified mereotopology of observable systems significantly into question.  The present paper shows that, at least within the ``consistent histories'' formalism of Griffiths, Omn\`es, Gell-Mann and Hartle (reviewed by \cite{omnes:92, griffiths:02}; see \cite{schloss:04, schloss:07} for comparison with other approaches), a non-trivial mereotopology of informally-identifiable systems cannot be defined self-consistently.  It shows, in particular, that macroscopic ``parts'' of the universe - ordinary things such as laboratory apparatus - cannot be consistently represented by TPSs of $\mathcal{H}_{\mathbf{U}}$.  It thus suggests that not only the expectation that a unique classical world dynamically emerges from quantum theory, but also the expectation that \textit{any} classical world dynamically emerges from quantum theory may be misguided.

\section{Consistent histories}

The formalism of consistent histories developed in \cite{griffiths:02} Ch. 8 and 10 is employed here.  Consider a Hilbert space $\mathcal{H}$ and a von Neumann projection $\lbrace \Pi_{i} \rbrace$ defined on $\mathcal{H}$.  By definition, the elements of $\lbrace \Pi_{i} \rbrace$ sum to the Identity on $\mathcal{H}$ and are mutually orthogonal; hence each component $\Pi_{j} = |j\rangle \langle j|$ for some complete, orthonormal basis $\lbrace |i\rangle \rbrace$ of $\mathcal{H}$.  A finite sequence of real measurement outcome values $\alpha_{1}, \alpha_{2}, ..., \alpha_{n}$ obtained by acting with $\lbrace \Pi_{i} \rbrace$ at a sequence of distinct times $t_{1}, t_{2}, ..., t_{n}$ can, therefore, be represented as a \textit{history} operator $Y$ defined by:

\begin{equation}
Y(\breve{\mathcal{H}}) = (\Pi_{1} \odot \Pi_{2} \odot ... \odot \Pi_{n})(\mathcal{H}_{\mathrm{1}} \odot \mathcal{H}_{\mathrm{2}} \odot ... \odot \mathcal{H}_{\mathit{n}}), \label{hist}
\end{equation}

where the index ranges over observation times and as in Eqn. 8.5 of \cite{griffiths:02} `$\odot$' represents the tensor product operating on entities indexed to different times and the $\mathcal{H}_{\mathit{i}}$ are time-indexed copies of $\mathcal{H}$.

If ``minimal'' quantum mechanics - quantum mechanics with no ``collapse of the wave function'' - is assumed as all available experimental evidence indicates it can be \citep{schloss:06}, then the underlying physical dynamics described by any such history must satisfy the Schr\"odinger equation.  This is assured by requiring that the physical dynamics separating any two observation times $t_{j}$ and $t_{k}$ are unitary, i.e. that the time propagation operator $T(t_{j},t_{k})$ is such that:

\begin{equation}
\Pi_{j} = T(t_{j},t_{k}) \Pi_{k} T(t_{k},t_{j}).
\end{equation}

For such histories, the \textit{chain operator} $K$ as defined by:

\begin{equation}
K(\Pi_{j} \odot \Pi_{k}) = \Pi_{k} T(t_{k},t_{j}) \Pi_{j}
\end{equation}

is such that the Born-rule probability $W(Y)$ of a history $Y$ is given by:

\begin{equation}
W(Y) = Tr(K^{\dag}(Y)K(Y)),
\end{equation}

where `$Tr$' is the operator trace (cf. \cite{griffiths:02} Eqn. 10.2, 10.3).  If the $\Pi_{i}$ composing $Y$ are all orthogonal, $K(Y)$ is the transition amplitude from the initial state of $Y$ to the final state; $W(Y)$ is the square of this amplitude.

The superposition principle assures that any final state $\psi_{f} \in \mathcal{H}$ can be reached from any initial state $\psi_{i} \in \mathcal{H}$ by multiple distinct sequences of physical state transitions.  Let  $Y^{\mu}$ and $Y^{\nu}$ be histories representing two such sequences; these histories have Born-rule weights $W(Y^{\mu})$ and $W(Y^{\nu})$ respectively.  \textit{Consistency} requires that the Born-rule weights behave as classical transition probabilities, i.e. that they be additive.  This condition is achieved provided the histories $Y^{\mu}$ and $Y^{\nu}$ do not interfere, i.e. provided that:

\begin{equation}
Tr(K^{\dag}(Y^{\mu})K(Y^{\nu})) = 0. \label{cons}
\end{equation}

Histories from a given $\psi_{i}$ to a given $\psi_{f}$ that do not interfere are also called ``mutually decoherent''; a collection of such histories is a \textit{framework}.  Griffiths emphasizes that frameworks, because they are defined relative to the unitary dynamics represented by the propagators $T(t_{k},t_{j})$, encompass \textit{all} interactions that affect a system; in particular, the time evolution of any system $\mathbf{S}$ interacting with its environment $\mathbf{E}$ can only be consistently described within a framework encompassing $\mathbf{S \otimes E}$ (\cite{griffiths:02} p. 123).

Frameworks based on mutually non-commuting measurement operators, such as position and momentum, are \textit{incompatible}; within the consistent-histories formulation of quantum mechanics, any statements involving conjunctions of outcomes from incompatible frameworks are regarded as \textit{meaningless}.  For example, it is meaningless to ask of a single particle traversing a double-slit apparatus both which path it took through the slits and where it landed on the screen.  More generally, any extraction of classical information from a system that does not commute with the physical dynamics specified by a given framework is incompatible with that framework.   

\section{Mereological partitions} 

A mereology on some domain $\mathbf{D}$ is a formal theory intended to capture the intuitive notion that some element $x_{j}$ of $\mathbf{D}$ can be a part of some other element $x_{k}$ of $\mathbf{D}$, for example, the intuitive notion that a vacuum chamber or an ion source can be part of an ion-trap apparatus.   Let `$x_{j} \leq x_{k}$' indicate this parthood relation.  An extensional mereology is one in which two entities that share all the same parts are identical.  A mereotopology is an extension of a mereology that formalizes the notions that entities can have interiors and boundaries, and that entities can be connected to or touching each other \citep{varzi:94, varzi:96, smith:96, casati:99}; a mereotopology enables the formalization of such statements as ``the ion source is within the interior of the vacuum chamber'' or ``the vacuum pump is connected to the vacuum chamber.''  The notion that entities either have or are contained within boundaries can itself be extended by formalizing the idea that boundaries partition the domain $\mathbf{D}$ into multiple interiors, with respect to any of which the remainder of $\mathbf{D}$ is exterior.  In the notation of \citet{smith:02}, a \textit{partition} $A$ over $\mathbf{D}$ is a collection of \textit{cells} that together cover $\mathbf{D}$.  Cells may be simple, or may be refined into collections of sub-cells; `$z_{j} \leq_{A} z_{k}$' will indicate this sub-cell relation in $A$.  It is explicitly assumed that the total number of cells in $A$ is finite; hence all refinements are finite and terminate in simple cells.  Intuitively, a partition is overlaid on a domain in the way a coordinate grid is overlaid on a map: an object $x \in \mathbf{D}$ may be \textit{located} in a cell $z \in A$, in which case we write $L_{A}(x,z)$.  If an object is located in multiple cells, extensionality requires that the cells must overlap:

\begin{equation}
L_{A}(x,z_{j}) \wedge L_{A}(x,z_{k}) \Rightarrow \exists z_{i} \in A~ \text{such that} ~(z_{i} \leq_{A} z_{j} \wedge z_{i} \leq_{A} z_{k}). \label{pcr}
\end{equation}

\citet{smith:02} call \eqref{pcr} the ``Principle of Classical Realism'' as it forbids the simultaneous occupation of separate locations by ``real'' objects.

A partition $A$ is said to \textit{recognize} an object $x$ that has a location in $A$:

\begin{equation}
x \in A =_{def} \exists z \in A~ \text{such that} ~L_{A}(x,z).
\end{equation}

If $A$ is to preserve the native mereological structure of $\mathbf{D}$ - that is, if it to capture the full parthood structure that $\mathbf{D}$ actually displays - it must recognize the parts of a recognized object $x$ and must, moreover, locate them within a subcell of the location of $x$:

\begin{align}
\forall x_{j}, x_{k} \in \mathbf{D}, \mathit{[x_{j} \leq x_{k}} ~\wedge~ &\exists z_{k} \in A~ \text{such that} ~L_{A}(x_{k},z_{k}) \Rightarrow \nonumber \\
&\exists z_{j} \in A~ \text{such that} ~z_{j} \leq_{A} z_{k} \wedge L_{A}(x_{j},z_{j})]. \label{dist}
\end{align}

However, if $z_{k}$ is a simple cell of $A$, in which case $z_{j} = z_{k}$, $A$ cannot be considered to recognize $x_{j}$ as distinct from $x_{k}$; in this case the mereology provided by $A$ is weaker than the native mereology of $\mathbf{D}$.  Let us, therefore, define for $x \in \mathbf{D}$ and $z \in A$ a notion of \textit{strong location} with the conditions:

\begin{equation}
SL_{A}(x,z) =_{def} L_{A}(x,z) ~\wedge~ \forall y \in \mathbf{D}, \mathit{L_{A}(y,z) \Rightarrow y \leq x}, \label{sl}
\end{equation}

and a notion of \textit{strong recognition} as:

\begin{equation}
x \in_{S} A =_{def} \exists z~ \text{such that} ~SL_{A}(x,z). \label{sr}
\end{equation}

Intuitively, condition \eqref{sr} requires strongly recognized things to have distinct locations; if multiple distinct parts of some object $x$ - e.g. multiple atoms - that have equivalent granularity are located in the same cell of $A$, they are not distinguishable by their location and hence are not strongly recognized.  A \textit{strong partition} on $\mathbf{D}$ can now be defined as a partition in which every cell is either empty or strongly locates some object in $\mathbf{D}$.  If $A$ is a strong partition on $\mathbf{D}$, $x \in \mathbf{D}$, $z \in A$ and $SL(x,z)$, the boundary of $z$ in $A$ can be regarded as a \textit{de facto} boundary of $x$ in $\mathbf{D}$, and $A$ can be regarded as providing a complete mereotopology of $\mathbf{D}$, i.e. one that fully captures both its native mereological properties and its boundaries.  It is this ability to employ strong partitions as mereotopologies on ``real'' domains that renders them valuable as formal tools.

The use of strong partitions is commonplace in physics, for example when a volume of fluid identified by its coordinates in some initial state is tracked through subsequent state changes by employing the initial-state coordinate ``box'' as a smoothly-deformable, transparent physical boundary that permits interactions but prevents the flow of matter.  Intuitively, strong partitions are minimal partitions at a given level of granularity; a non-strong partition simply adds cells that are superfluous from the point of view of information about object locations.  If a domain $\mathbf{D}$ does not admit a strong partition, it does not admit any partition that is mereologically significant; such domains either contain no objects, or contain objects that can only be organized as sets.  In what follows, therefore, all mentioned partitions will be taken to be strong partitions.

Motivated explicitly by the notion of history defined in \S 2 above, \cite{smith:02} define a history as a time indexed sequence of partitions $A_{i}$.  Such a history is consistent in case all of the time-indexed statements $L_{i}(x_{j},z_{k})$ describing the locations of recognized objects are mutually consistent; this will be the case provided they are consistent for each time $t_{i}$, i.e. provided no recognized object is required to be in two separated cells simultaneously.  A \textit{library} of histories is a collection of mutually exclusive histories that is maximal in the sense that probabilities can be assigned to the histories so as to sum to unity.  This notion of history specifies no requirements regarding the dynamics that transform a particular statement $L_{i}(x,z_{i})$ into its successor $L_{i+1}(x,z_{i+1})$.  While \citet{smith:02} briefly discuss the formal analogy between histories of partitions and consistent histories as defined within quantum mechanics and note that quantum systems ``can ... contravene the Principle of Classical Realism as formulated'' by \eqref{pcr}, they do not examine the conditions under which quantum systems violate \eqref{pcr} or attempt to overlay a history of partitions onto unitary dynamics in Hilbert space.

\section{Overlaying partitions on Hilbert space}

We can now ask: can a history of partitions be overlaid onto Hilbert space, and in particular, onto the universal Hilbert space $\mathcal{H}_{\mathbf{U}}$, in a way that (i) provides a \textit{de facto} mereotopology of ``systems'' and (ii) satisfies the consistency conditions given by \eqref{cons}?  As a ``system'' can, in principle, comprise any collection of physical degrees of freedom, it would appear that (i) can be satisfied trivially by regarding any system $\mathbf{S}$ as defined by the Hilbert space spanned by the basis vectors of $\mathcal{H}_{\mathbf{U}}$ corresponding to its particular degrees of freedom.  As noted earlier, this trivial solution for (i) is being employed whenever someone implicitly specifies a Hilbert space $\mathcal{H}_{\mathbf{S}}$ by specifying a state $|\mathbf{S}\rangle$ as a linear combination of basis vectors $\mathit{\sum_{i} \lambda_{i} |s_{i}\rangle}$.  Our question is, therefore, twofold: does the trivial solution to (i) satisfy \eqref{cons}, and if not, can any physically-reasonable conditions be added to (i) that will make it satisfy \eqref{cons}?  The first of these questions is easily answered: it is the failure of (i) to assure consistency that motivates the consistent-histories formulation of quantum mechanics.  What must be added to (i) is known: it is the notion of a framework, with the concomitant rejection as meaningless of any statements not consistent with the adopted framework.  The issue to be settled, then, is whether \eqref{pcr} can be made compatible with \eqref{cons} in the context of a fixed framework.

Recall from \S 2 that a framework is a collection of mutually-noninterferring histories, and from \eqref{hist} that the basic component of a history is the action at some time $t$ of a specific projector $\Pi_{j}$ on the Hilbert space $\mathcal{H}_{\mathbf{S}}$ representing some system $\mathbf{S}$.  Any finite physical implementation of a projection $\lbrace \Pi_{i} \rbrace$ has a minimum resolution $\epsilon$ below which results cannot be distinguished; in particular, the number $m_{\epsilon}$ of component projectors $\Pi_{j} \in \lbrace \Pi_{i} \rbrace$ for which $\langle \mathbf{S} |\mathit{\Pi_{j}}| \mathbf{S}\rangle \geq \epsilon$ and hence for which the outcome value $\alpha_{j}$ associated with $\Pi_{j}$ can be recorded as an observation is finite.  At the $k^{th}$ measurement time in a history $Y = \Pi_{1} \odot \Pi_{2} \odot ... \odot \Pi_{n}$ indexed by measurement times, therefore, the projector $\Pi_{k}$ that is included in the history must be unique, and must be one of the $m_{\epsilon}$ components of $\lbrace \Pi_{i} \rbrace$ for which $\langle \mathbf{S} |\mathit{\Pi_{j}}| \mathbf{S}\rangle \geq \epsilon$.  Each time point $t_{k}$ of a history $Y$ can, in other words, be considered as a mapping:  

\begin{equation}
t_{k}^{Y} : \lbrace \Pi_{i} \rbrace \rightarrow \lbrace \mathrm{1, 2}, ..., \mathit{m_{\epsilon}} \rbrace
\label{tpart1}
\end{equation}

that selects one of the $m_{\epsilon}$ possible projectors $\Pi_{k}$.  This mapping satisfies \eqref{pcr} and is hence a partition $A_{k}$ at $t_{k}$; a sequence of such mappings at a sequence of times is, therefore, a history of partitions.

The partition expressed by \eqref{tpart1} is a partition of the space of possible outcomes, not a partition of $\mathcal{H}_{\mathbf{S}}$ as desired.  To obtain a partition of $\mathcal{H}_{\mathbf{S}}$, note that the images in $\mathcal{H}_{\mathbf{S}}$ of the $\Pi_{j}$ are forced to be distinct by the orthogonality of the $\Pi_{j}$.  Hence each time point $t_{k}$ of a history $Y$ can also be regarded as a mapping:

\begin{equation}
t_{k}^{Y} : \mathcal{H}_{\mathbf{S}} \rightarrow \lbrace \mathrm{1, 2}, ..., \mathit{m_{\epsilon}} \rbrace
\label{tpart2}
\end{equation}

where here the integers $1, 2, ... m_{\epsilon}$ are regarded as indexing the $m_{\epsilon}$ images of the projectors that can act on $\mathcal{H}_{\mathbf{S}}$ at $t_{k}$.  The orthogonality of the $\Pi_{j}$ assures that this mapping satisfies \eqref{pcr} and is hence a partition at $t_{k}$.  A time series of such mappings is, therefore, a history of partitions of $\mathcal{H}_{\mathbf{S}}$ as desired. 

This procedure is easily extended to a ``universe'' comprising $N$ non-interacting and hence mutually separable (i.e. unentangled) ``systems'' $\mathbf{S}_{\mathit{\mu}}$.  In this case $\mathcal{H}_{\mathit{i}} = \otimes_{\mu} \mathcal{H}_{\mathit{\mu} \mathit{i}}$ and a projection acting on $\mathbf{S}_{\mathit{\mu}}$ can be represented as $\lbrace \Pi^{\mu}_{j} \otimes (\otimes_{\nu \neq \mu} I^{\nu}) \rbrace$ where $I^{\nu}$ is the identity on $\mathcal{H}_{\mathit{\nu}}$; these operators clearly mutually commute.  Similarly the propagation operator for $\mathbf{S}_{\mathit{\mu}}$ is $T^{\mu}(t_{i},t_{i+1}) \otimes (\otimes_{\nu \neq \mu} I^{\nu})$; these operators also mutually commute.  Here, as in the case above, each measurement time $t_{k}^{\mu}$ selects a unique projector $\Pi^{\mu}_{k}$ independently for each of the component systems.  The selected projector indices $k^{\mu}$ define mutually-exclusive simple cells of a partition $A_{k}$ and a history of partitions can be constructed.

From both mereological and physical perspectives, however, the histories constructed above are trivial: the partition of $\mathcal{H}_{\mathbf{S}}$ defined by \eqref{tpart2} does not describe a parthood relation, while its extension to $\otimes_{\mu} \mathcal{H}_{\mathit{\mu}}$ specifically disallows physical interactions between the ``parts.''  Constructing a mereologically and physically non-trivial history requires relaxing this latter constraint.  Consider, therefore, a universe $\mathbf{U}$ characterized by a large number of interacting degrees of freedom $\psi, \varphi, \chi, ...$ and consider a partition $A_{i}$ at $t_{i}$ of this universe into two interacting components, a ``system'' $\mathbf{S}$ characterized by more than one degree of freedom and its environment $\mathbf{E}$ as given by the TPS $\mathcal{H}_{\mathbf{U}} = \mathcal{H}_{\mathbf{S}} \otimes \mathcal{H}_{\mathbf{E}}$.  A consistent history $..., A_{j}, A_{k}, A_{l}, ...$ of such partitions that satisfied \eqref{pcr} would confer intuitive ``objecthood'' on $\mathbf{S}$ by sequestering its degrees of freedom in a sequence of cells $..., z_{j}, z_{k}, z_{l}, ...$ and hence distinguishing them from the degrees of freedom of $\mathbf{E}$, which would be contained within cells other than the $z_{i}$.  Our question becomes, does such a history of (strong) partitions that is consistent in the sense of satisfying \eqref{cons} exist?

Approaching this question requires asking how the function $SL_{A_{i}}(x,z)$, the function that \textit{identifies} an entity $x$ as being inside a cell $z$ of $A_{i}$ at $t_{i}$, is physically implemented.  In the case of interest, ``entities'' are degrees of freedom and only two cells need be considered: one corresponding to $\mathbf{S}$ and the other to $\mathbf{E}$.  That some degree of freedom $\varphi$ is contained within the cell $z_{\mathbf{S}}$ corresponding to $\mathbf{S}$ at time $t_{i}$ is an item of classical information.  Quantum theory provides only one way of extracting classical information from quantum degrees of freedom: action on those degrees of freedom with a projection.  Call this projection $\lbrace \Pi^{A_{i}}_{\mathbf{S}}, \Pi^{A_{i}}_{\mathbf{E}} \rbrace$ at $t_{i}$, and the history it induces $Y^{A}$.  From \eqref{tpart2}, each time point of $Y^A$ can be considered as a mapping:

\begin{equation}
t_{k}^{Y^{A}} : \mathcal{H}_{\mathbf{U}} \rightarrow \lbrace \mathit{z}_{\mathbf{S}}, \mathit{z}_{\mathbf{E}} \rbrace
\label{tpart3}
\end{equation}

where for any degree of freedom $\varphi$ of $\mathbf{U}$,

\begin{equation}
t_{k}^{Y^{A}} : \varphi \mapsto \mathit{z}_{\mathbf{S}} ~\text{or}~ t_{k}^{Y^{A}} : \varphi \mapsto \mathit{z}_{\mathbf{E}}, \label{sep}
\end{equation}

but to preserve consistency with \eqref{pcr}, no degree of freedom maps to both $\mathit{z}_{\mathbf{S}}$ and $\mathit{z}_{\mathbf{E}}$.  A history $Y^{A}$ of (strong) partitions $A_{i}$ exists for $\mathcal{H}_{\mathbf{U}}$ just in case a projection satisfying both both \eqref{cons} and \eqref{sep} exists.

Having effectively translated \eqref{pcr} into a constraint on the actions of individual projectors on individual degrees of freedom, we can now test \eqref{cons} directly by asking whether histories generated by $\Pi^{A_{i}}_{\mathbf{S}}$ and $\Pi^{A_{i}}_{\mathbf{E}}$ acting separately are decoherent, i.e. whether the actions of $\Pi^{A_{i}}_{\mathbf{S}}$ and $\Pi^{A_{i}}_{\mathbf{E}}$ interfere for any chosen degrees of freedom $\varphi$ and $\chi$ or whether:  

\begin{equation}
\Pi^{A_{i}}_{\mathbf{S}} \Pi^{A_{i}}_{\mathbf{E}} |\varphi \chi\rangle = \Pi^{A_{i}}_{\mathbf{E}} \Pi^{A_{i}}_{\mathbf{S}} |\varphi \chi\rangle. \label{coms}
\end{equation}

The condition under which \eqref{coms} is satisfied is well known: it is the separability condition $|\varphi \chi\rangle = |\varphi\rangle \otimes |\chi\rangle$ that assures that the composite state $|\varphi \chi\rangle$ is not entangled.  Hence $\mathbf{S}$ can be separated from $\mathbf{E}$ by the partition $A_{i}$ only if none of their respective degrees of freedom are entangled.  This separability condition is satisfied in general, however, only if none of their respective degrees of freedom interact: any $\varphi - \chi$ interaction representable by a unitary propagator, and hence any interaction that satisfies the Schr\"odinger equation, will entangle the composite state $| \varphi \chi \rangle$.  A universe $\mathbf{U}$ can, therefore, be partitioned into mutually exclusive collections of degrees of freedom describing distinct ``objects'' $\mathbf{S}$ and $\mathbf{E}$ only if these objects do not interact.

This result generalizes from projective measurements of $\mathbf{S}$ to generalized measurements that employ positive operator-valued measures (POVM), collections of positive semi-definite operators on a Hilbert space $\mathcal{H}$ that sum to the identity on $\mathcal{H}$ but are not required to be orthogonal \citep{nielsen-chaung:00}.  A POVM $\lbrace E_{i} \rbrace$ acting on $\mathcal{H}_{\mathbf{S}}$ can be represented as a projection $\lbrace \Pi_{i} \rbrace$ acting a notional ancilla with the dimensionality of $\lbrace \Pi_{i} \rbrace$ that is associated with $\mathcal{H}_{\mathbf{S}}$.  Such an ancilla is a collection of degrees of freedom of $\mathbf{E}$; it is commonly interpreted as an ``apparatus'' that interacts with $\mathbf{S}$, the state of which is determined by the projective measurement $\lbrace \Pi_{i} \rbrace$.  The result also generalizes to ``mixed'' states of $\mathbf{S}$, i.e. states of an ensemble of copies of $\mathbf{s}$, as it applies to the interaction between each element of the ensemble and its surrounding environment.

It is often suggested, under the rubric of ``the emergence of classicality'' (e.g. \cite{schloss:04, schloss:07, landsman:07, wallace:08, hartle:08, omnes:08, griffiths:11}), that  system-environment decoherence provides approximate separability between observable systems that is ``good enough'' to regard them as partitionable objects.  System-environment decoherence, however, \textit{requires} system-environment entanglement.  Hence effective separability driven by decoherence fails not because individual ``objects'' cannot be separated from each other, but because such ``objects'' cannot be separated from their mutual environment (cf. \cite{fields:12b} where this conclusion is reached by different means).  This failure is masked, in most discussions of emergence, by an implicit and \textit{a priori} assumption that stipulated system-environment boundaries are somehow preserved by the universal Hamiltonian $H_{\mathbf{U}}$, an assumption for which a physical justification is difficult to imagine.

\section{Conclusion}

The assumption that ``systems'' manipulated in the laboratory have well-defined Hilbert spaces is convenient, and from the perspective of practical calculations, probably essential.  It is, therefore, made ubiquitously in discussions of the foundations of quantum mechanics.  Previous work has questioned this assumption from a variety of perspectives \citep{fields:10, fields:11, fields:12a, fields:12b, fields:12c}.   It is shown here that a non-trivial mereological partition can be imposed on a quantum universe only if that universe is assumed to be free of interactions, i.e. to be physically trivial.  This result provides additional support for the view that ``systems'' cannot be regarded as foundational entities in quantum theory.

\end{document}